\documentstyle[twocolumn,amssymb,aps,psfig,floats]{revtex}
\begin{document}
\draft

\twocolumn[\hsize\textwidth\columnwidth\hsize\csname @twocolumnfalse\endcsname

\title{Dynamical properties of the compounds 
CuGeO$_3$ and $\alpha'$--NaV$_2$O$_5$ \newline
at temperatures $T\neq 0$.}

\author{K.~Fabricius}
\address{Physics Department, University of Wuppertal, 42097 Wuppertal}
\author{U.~L\"ow}
\address{Physikalisches Institut der Johann Wolfgang Goethe Universit\"at,
D-60054 Frankfurt am Main, Germany}

\date{\today}
\maketitle

\begin{abstract}
\begin{center}
\parbox{14cm}{
The dynamical structure factor of the frustrated spin 
1/2 Heisenberg model 
for a series of frustration parameters 
at finite temperatures is presented.
A sharp upper boundary of the spinon continuum is found. A simple
method to extract the nearest neighbour coupling $J$ from the
spectral width is suggested.
Signatures of frustration are discussed
and a comparison with neutron inelastic scattering data 
is given.
}
\end{center}
\end{abstract}

\pacs{
\hspace{1.9cm}
}
\vskip2pc]

The recent widespread interest in low-dimensional 
quantum spin systems has been brought up and maintained 
by the discovery of new compounds with a variety of 
spin-spin couplings. Particular effort has been  
concentrated on the first two inorganic substances
undergoing a Spin-Peierls transition:
CuGeO$_3$ and $\alpha'$--NaV$_2$O$_5$. 

There is extensive literature on the low temperature properties 
of CuGeO$_3$ (see for example Ref.\cite{APH97} and references therein),
 but there is still no agreement on the model describing 
the uniform phase ( $T>T_{SP}$ ) also.
If the interchain interaction is negligible
the most likely 
candidate is the isotropic frustrated spin 1/2 Heisenberg model
\cite{CCE,RD}
in one dimension

\begin{equation}
H = 2 J \sum_{i = 1}^{N} \left\{ \vec S_{i} \cdot \vec S_{i + 1}  
+ \alpha \vec S_{i} \cdot \vec S_{i + 2}  
\right\}  
$$
\label{Hfrus}.
\end{equation}
\noindent

Alternatively Uhrig proposed a model including interchain interaction with
a small value of inchain next to nearest neighbour coupling 
$\alpha$ \cite{uhrig}.

In a recent study \cite{us2} it has been demonstrated compellingly
that in the framework of the first model (Eq. (\ref{Hfrus}))
for the unique set of parameters $J/k_B=80K$ and  $\alpha =0.35$
excellent agreement between experiment and
theory is achieved 
for the magnetic susceptibility. 

In the following we focus on the dynamical properties of both substances 
at temperatures  $k_B T\geq 0.6 J$. 
We compare our theoretical results with data
obtained from neutron inelastic scattering (NIS) experiments \cite{AFM}
performed on CuGeO$_3$. There are still no corresponding results for 
$\alpha'$--NaV$_2$O$_5$.
If it is confirmed that $\alpha'$--NaV$_2$O$_5$ is well described by the pure 
isotropic Heisenberg model our results for $\alpha =0$ can be taken as a 
prediction for this compound.

\begin{figure}[hbt]
\begin{center}
\mbox{\psfig{figure=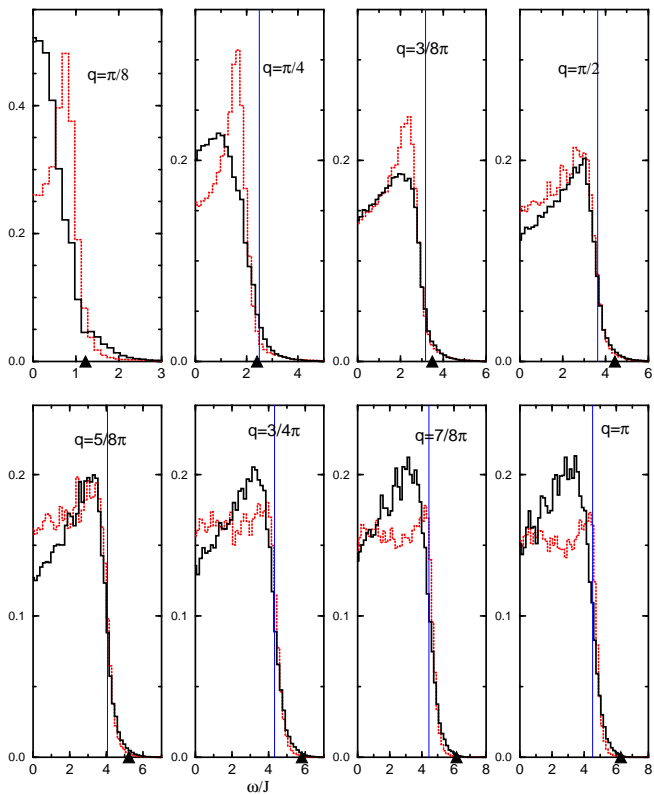,width=8cm,angle=0}}
\end{center}
\caption{ Dynamical structure factor 
for $\alpha=0.0$ (black solid line) and $\alpha=0.35$ (red dashed line), 
as a function of $\omega/J$ for $k_B T=3.8J$. The triangles mark the upper boundary 
of the two-spinon continuum for the XXX-model. The blue vertical lines mark 
the experimental spectral boundaries read off 
from Fig.4 in Ref. \protect\cite{AFM}. }
\label{fig1}
\end{figure}

\begin{figure}[hbt]
\begin{center}
\mbox{\psfig{figure=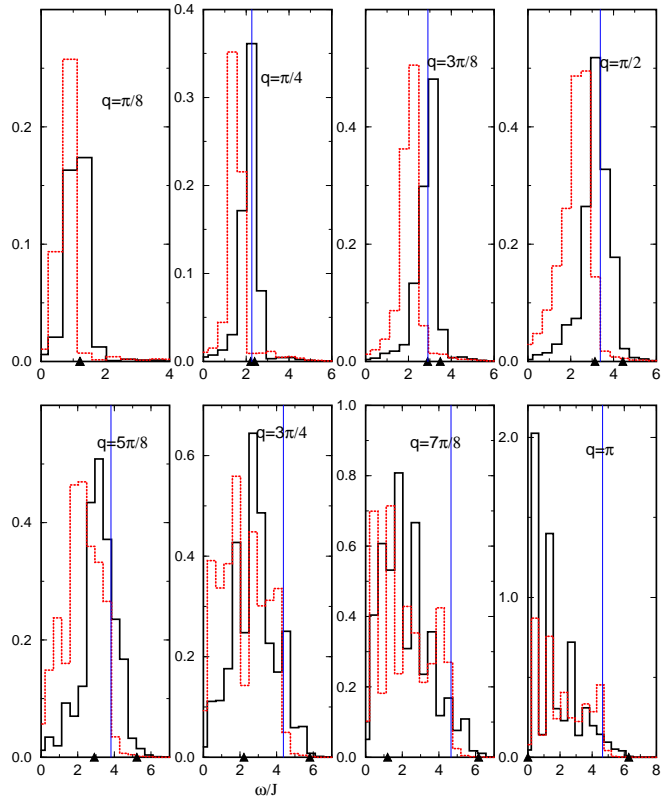,width=8cm,angle=0}}
\end{center}
\caption{ As Fig.\protect\ref{fig1} but for $k_B T =0.6J$. 
The triangles mark the upper
and lower boundaries of the two-spinon continuum for the XXX-model. The blue vertical lines mark 
the experimental spectral boundaries read off 
from Fig.1b in Ref. \protect\cite{AFM}.}
\label{fig2}
\end{figure}

We have determined the dynamic structure 
factor $S(q,\omega)$ at finite $T$ for the frustrated Heisenberg model
and the $XXX$-model 
for chains with $N=  16, 18$ spins and periodic boundary 
conditions by the method of complete and exact diagonalization (CED)
as described in detail in Ref. \cite{us}. This was done for
$\alpha = 0.0, 0.06, 0.12, 0.18, 0.24, 0.35, 0.45, 0.5 $.
The largest subspaces to be treated are of dimension 415 for
$N=16$ and 1367 for $N=18$ (see also  Ref. \cite{SSS}).
Our aim is to study  the dependence of $S(q,\omega)$ on the frustration
parameter $\alpha$ at $T \neq 0$ and to highlight the unambiguous features 
distinguishing models with varying degree of frustration.

We study chains with maximal 18 spins. 
Lanczos studies of ground state properties of spin $1\over 2$ systems 
routinely handle chains of 30 spins. So one could dismiss the present work
 as not worthwhile
because it is expected to be too inaccurate. But 
thermodynamical observables can be determined by CED with high accuracy
for temperatures $ {k_B T} \geq 0.6 J $. 
An extensive investigation of finite size errors of dynamical correlations
convinced us that this method, 
if carefully applied, gives sufficiently accurate
results with easily controllable errors, which are not obscured by more
or less arbitrary approximations.
The obvious reason is that 
the number of contributing matrix elements is for $T>0$ many orders of
magnitude larger than for $T=0$.
We refer to Ref. \cite{us} 
where a thorough finite size analysis is presented and 
to remarks in the course of this paper.
With substantially more effort marginally longer chains could be 
studied exactly. But probably the additional accuracy would barely justify 
the effort. The real challenge is instead to extract as much reliable
results as possible from the existing data.

\begin{figure}[hbt]
\begin{center}
\mbox{\psfig{figure=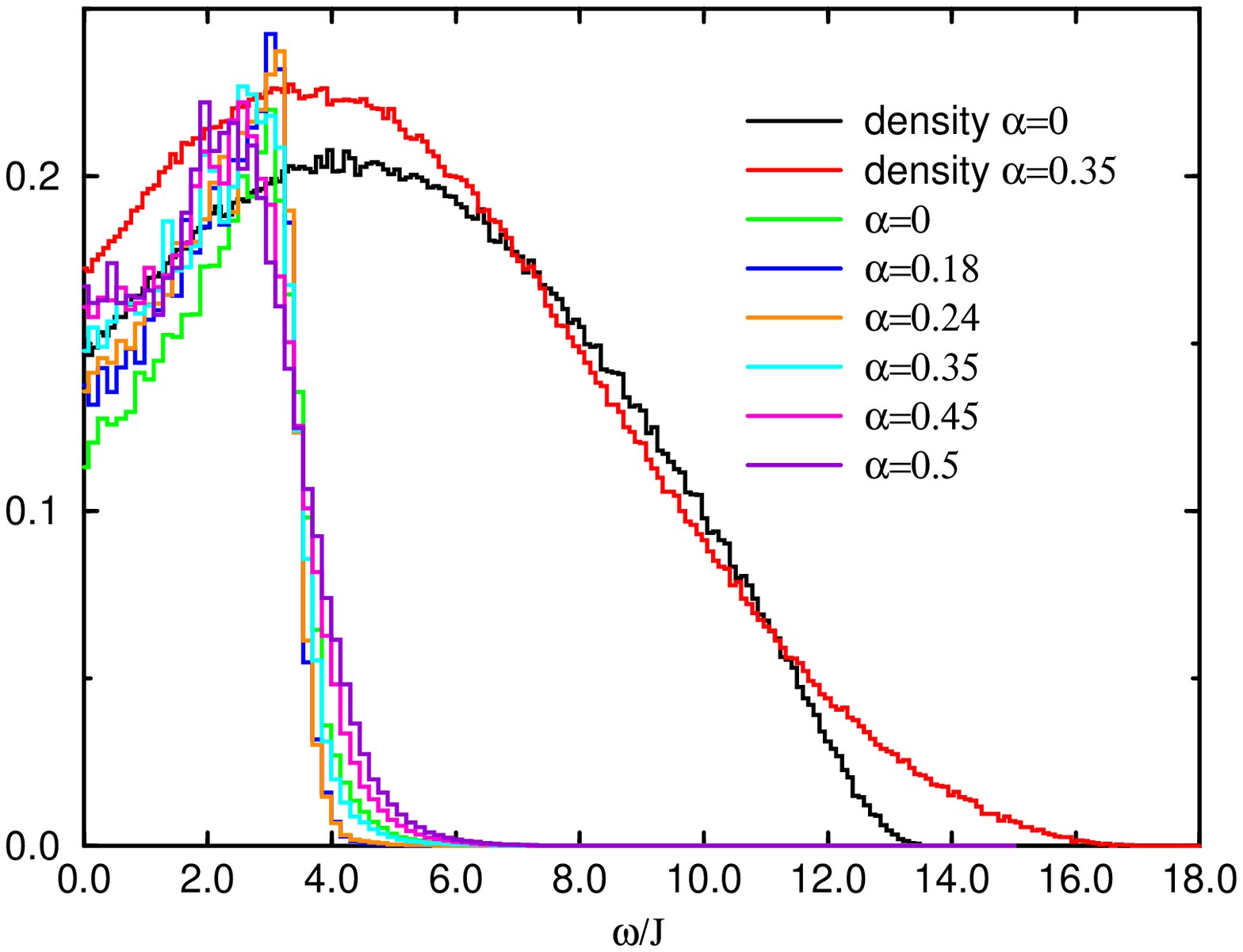,width=7cm,angle=0}}
\end{center}
\caption{$S(q=\pi/2,\omega)$ for  $\alpha=0.0, 0.18, 0.24, 0.35, 0.45, 0.5 $ 
at $k_B T = 3.2 J$ and the density of contributing excitations for $\alpha=0$
and $\alpha=0.35$ (rescaled).}
\label{fig3}
\end{figure}

\begin{figure}[hbt]
\begin{center}
\mbox{\psfig{figure=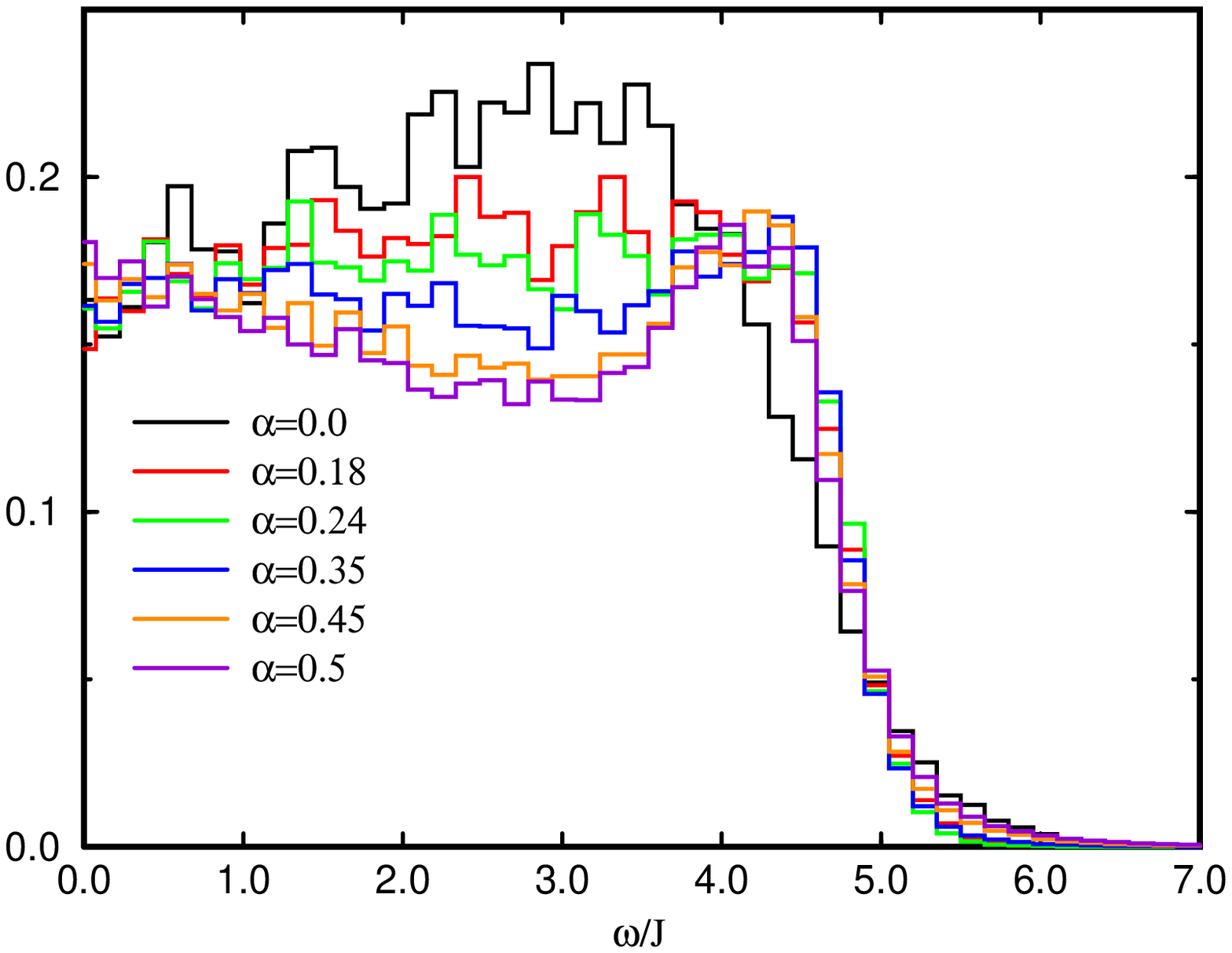,width=7cm,angle=0}}
\end{center}
\caption{$S(q=\pi,\omega)$ for $\alpha =0.0, 0.18, 0.24, 0.35, 0.45, 0.50$
and $k_B T=3.2J$.}
\label{fig4}
\end{figure}

For $k_B T< 0.6 J $ the method applied by us gives only poor results. We
therefore restricted ourselves to temperatures $k_B T \geq 0.6 J$
and considered qualitative and global features respectively. 
For higher temperatures where our
method is accurate 
the existing experimental results are of qualitative
nature only and our data can serve 
in the future to disentangle magnetic and phononic contributions
in the analysis of experimental data.

We find that the 
structure function has for all practical purposes (but not strictly)
a bounded support in the $q-\omega$ plane. 
The lower boundary - well known from spin dynamics at $T=0$ -
disappears gradually for growing $T$ (see Figs.\ref{fig1}, \ref{fig2}), 
whereas the upper boundary is nearly
unchanged up to infinite temperature. The structure function develops
for increasing $T$ a steep slope at its boundary. We emphasize that
the boundary as well as the functional form
of $S(q,\omega)$ near the boundary 
determined by CED show very small
finite size errors: While the precise positions of local peaks of $S(q,\omega)$
for small and intermediate $\omega$ are not perfectly $N$-independent 
for small $T$,
the shape near the boundary and the location of the boundary are remarkably
stable. In any case the spectral boundary is easily determined independently
of the criteria applied.
This is in sharp contrast to the situation at $T=0$ where the standard
tool to study spin dynamics is the recursion method \cite{VM},
which works well for low lying excitations but is much less accurate
for higher ones.

It is a remarkable and unexpected result that for  $T>0$ the spectral density
is bounded by a curve similar (but not equal) to the two spinon boundary
of $S(q,\omega)$ in the $XXX$-model detected by M\"uller et al. \cite{MTPB81} 
at $T=0$. For $T \geq 0$ this was first observed in Ref. \cite{us}.

As for $ T=0 $  this results from a suppression of the
matrix elements 
$\left|\langle \nu|S_3(q)|\mu\rangle \right| $ 
for large energy transfer. The density of allowed pairs of
excitations  however is not responsible for the steep drop of
$S(q,\omega)$ near the boundary since it is still sizable for values of
the energy transfer far beyond the boundary of the two-spinon 
continuum. This is shown in Fig. \ref{fig3} for $\alpha = 0.0, 0.35$.

The spinon is a rigorously defined notion only in the context of
exactly solvable models (ESM) where Bethe-Ansatz techniques work 
\cite{Yam69,JKM73,MTBB,FaTa81,Hal88,Sha88,BCK96}. 
It is indispensable to understand the physics of spin dynamics at $T=0$
as first noted and discussed exhaustively in 
\cite{MTBB}, see also \cite{HaZi93,KBM96}.
As we have just demonstrated, certain aspects of spin dynamics at $T=0$
in ESM and $T>0$ ( up to infinite temperature ) 
even for not exactly solvable models are strikingly similar.
This makes it evident that the spinon is a much broader concept than 
originally anticipated.

Our most conspicuous new result is shown in
Figs. \ref{fig1}, \ref{fig2},\ref{fig4}: the spectral width
is essentially independent of $\alpha$ for $q=\pi$.
This is of eminent importance for the analysis of experiments:
it allows to determine the coupling $J$ and its temperature dependence
from neutron inelastic scattering
without any further assumption on the model parameters. 

\begin{figure}[hbt]
\begin{center}
\mbox{\psfig{figure=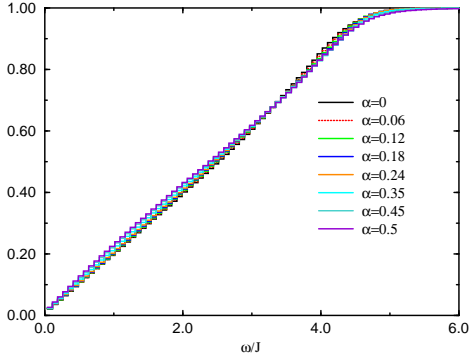,width=7cm,angle=0}}
\end{center}
\caption{ $\Sigma(q=\pi,\omega, T=\infty)$ for $N=16$ and 
$\alpha$=0.0, 0.06, 0.12,0.18, 0.24, 0.35, 0.45, 0.5.}

\label{fig5}
\end{figure}

A similar statement can be made about the normalized integrated intensity

\begin{equation}
\Sigma(q,\omega)=\frac{1}{S_\infty(q)} \int_0^{\omega} S(q,\omega ') d\omega' 
\end{equation}
\noindent
with $ S_\infty(q) =\int_0^{\infty} S(q,\omega ') d\omega'  $.
For $q=\pi$ there exists for each temperature an interval were these
functions are narrowly coalescing for all $\alpha$ values 
( see e.g. Fig. \ref{fig5}).

Arai et al. \cite{AFM} measured a spectral width of 32 meV at $T=50K$. This
gives $J/k_B=79K$ and is in perfect agreement with $J/k_B=80K$ determined 
from the
observation that the magnetic susceptibility has its maximum at $T=56K$ \cite{us2}.
So we observe that  completely independent experiments performed in
the same energy range give the same value for the coupling $J$. 
At $T=300K$ the measured width is 30 meV corresponding to $ J/k_B=68K$. 

Having determined $J$ by this very simple method, we obtain the
value of $\alpha$
from finer details of $S(q,\omega)$. These are:

(i) The theoretical width of the spectral density for $q\neq \pi$ shows for low
temperatures a distinct $\alpha$-dependence.
We extracted the boundary of the spectral density from the
color contour maps of Ref. \cite{AFM}. They are shown as vertical lines in 
Fig. \ref{fig1} and \ref{fig2}.
For $T=50K$ they coincide perfectly with the theoretical result valid for
$\alpha=0.35$. 

\begin{figure}[hbt]
\begin{center}
\mbox{\psfig{figure=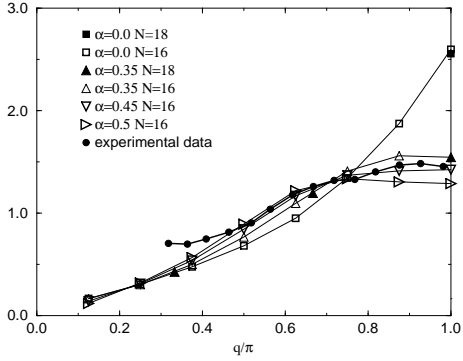,width=7cm,angle=0}}
\end{center}
\caption{ $S_\infty(q)$ at $k_B T=0.6 J$
for $\alpha=0.0, 0.35, 0.45, 0.5$ for $N=16,18$
and experimental data at $T=50 K$ read off from Fig.2 in Ref.\protect\cite{AFM}.
(The lower limit of integration for the theoretical curves
is 3meV  with $J/k_B=80K$ in accordance with Ref.\protect\cite{AFM}.)}
\label{fig6}
\end{figure}

(ii) The integrated intensity $S_\infty(q)$ 
is shown in Fig.\ref{fig6} for $T=50K$ in conjunction with data
from Arai et al. (see Fig. 2 of Ref.\cite {AFM}). If the vertical scale
of the experimental data 
is appropriately adjusted the agreement of experiment and theory is 
remarkable for $\alpha \approx 0.35$, whereas 
$\alpha =0 $ is clearly ruled out.
There is no question of finite-size errors, they are
much smaller than the experimental errors as shown in the figure.

(iii)  A further distinguishing qualitative feature is the trough-shaped 
structure detected experimentally which is fully developed at 
$T=300K$ (see Fig.4 of Ref.\cite{AFM}). 
Fig. \ref{fig4} shows that the same phenomenon is present for
$\alpha \geq 0.35 $ whereas for 
$\alpha =0$ we find a  hill-like structure.

Summarizing we have found an unexpected high degree of
correspondence of experimental data and theoretical results. We 
state that the model parameters  $J$ and $\alpha$ obtained from
thermodynamical properties of CuGeO$_3$ are also strongly 
favoured by NIS data.

 We conclude this section by predicting some properties of the
NIS cross-section of $\alpha'$--NaV$_2$O$_5$. Assuming $\alpha =0$
we find :

1.) The spectral boundary $\omega_B \pm \Delta \omega_{L,R}$ for  
$k_B T=1.2 J$ and $q=m \pi/8$
is given in the 
following table. Here $\Delta \omega_{R}$ is much larger than 
$\Delta \omega_L$, because $S(q,\omega)$ for $\alpha=0$ has a flat tail 
and becomes 
steeper for larger temperatures only.
\bigskip

\begin{tabular}{|c|c|c|c|}
\hline
\ \ m\ \ &\ \ \ $\omega_B/J$ \ \ & \ \ $\Delta\omega_L/J$ \ \ &
                                 \ \ $\Delta\omega_R/J$\ \ \\
\hline
1 &  2.04 &  0.16 &  0.60 \\
2 &  2.78 &  0.14 &  1.37 \\
3 &  3.84 &  0.43 &  1.33 \\
4 &  4.76 &  0.46 &  1.20 \\
5 &  5.22 &  0.48 &  0.77 \\
6 &  5.51 &  0.58 &  0.60 \\
7 &  5.95 &  0.44 &  0.32 \\
8 &  5.65 &  0.43 &  0.76 \\
\hline
\end{tabular}
\bigskip

2.) The integrated spectral density $S_\infty(q)$ at $k_B T=0.6 J$
is shown in Fig. \ref{fig6} for $\alpha =0$.

3.)  In clear distinction to CuGeO$_3$ the intensity is not suppressed
but enhanced around $\omega/J=3$ and $q=\pi$.

\begin{figure}[hbt]
\begin{center}
\mbox{\psfig{figure=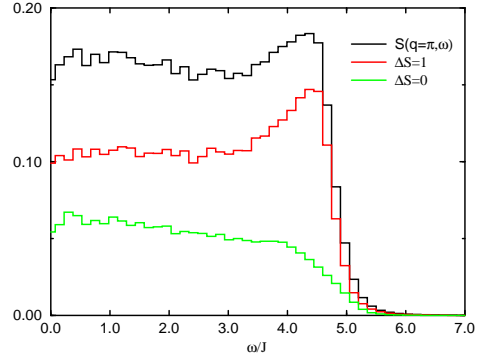,width=7cm,angle=0}}
\end{center}
\caption{$S(q=\pi,\omega)$  at $k_B T=3.2 J$ for 
$N=18$ and $\alpha=0.35$. Contribution from transitions
with
$\Delta S=0$ and $\Delta S=1$}
\label{fig7}
\end{figure}

For a discussion of the underlying physics we refer to Ref.\cite{us}.
We add the following supplement.
As we consider here solely isotropic models the spin $\vec S$ is conserved
and we can study another classification of excitations contributing to
$S(q,\omega)$. There are two classes of transitions behaving
markedly different:

1.) $\Delta S =0$ transitions:  They do not occur for $T=0$.
In Fig. \ref{fig7} it is shown that they contribute mainly to the central 
bulk of $S(q,\omega)$.

2.) $\Delta S =1$ transitions: Besides adding intensity to the central region
they build up the steep slope near the spectral boundary.

This note is exclusively devoted to an attempt to explain NIS data of
CuGeO$_3$ in the framework of model Eq. (\ref{Hfrus}). We postpone other
topics like a theoretically oriented study of spin dynamics
in frustrated models as well as the application to e.g.
Raman scattering to future publications.

\bigskip

We are grateful to M.Arai and M.Fujita for comments on their NIS data.
We thank A.~Fledderjohann, M.~Karbach and K.-H.~M\"utter for discussions
on spin dynamics at low temperatures.
U.L. gratefully thanks B.~L\"uthi for stimulating discussions and support
and M.~Braden for information on NIS.

\end{document}